  \documentclass[aps,twocolumn]{revtex4-1}
\usepackage{graphicx,amsmath}\usepackage{color}
\draft 

\begin{document}


\title{Generalized coherent states for position-dependent effective mass systems} 



\author{Naila Amir}

\email{naila.amir@live.com, naila.amir@seecs.nust.edu.pk}
\affiliation{School of Electrical Engineering and Computer Sciences, National University of Sciences and
             Technology, Islamabad, Pakistan}
\author{Shahid Iqbal}
\affiliation{School of Natural Sciences, National University of Sciences and
             Technology, Islamabad, Pakistan}
\email{sic80@hotmail.com, siqbal@sns.nust.edu.pk}


\date{\today}

\begin{abstract}

A generalized scheme for the construction of coherent states in the context of position-dependent effective mass systems has been presented. This formalism is based on the ladder operators and associated algebra of the system which are obtained using the concepts of supersymmetric quantum mechanics and the property of shape invariance. In order to exemplify the general results and to analyzed the properties of the coherent states, several examples have been considered.

\end{abstract}


\maketitle 

\section{Introduction}
Coherent states were initially introduced by Schr\"{o}dinger in 1926 \cite{c5} in the context of classical-quantum correspondence of dynamical systems. After a dormant period of more than three decades, these states were re-casted in 1963 by Glauber in the quantum mechanical description of coherent electromagnetic field \cite{G.1}. Due to their special properties, coherent states became indispensable to many research fields, such as, quantum optics \cite{gl.bk}, quantum information \cite{csqi1} and quantum computation \cite{csqc}. In particular their ability to get entangled in various optical systems arose them as to become a key resource in many modern technologies, such as, quantum meterology\cite{qm}, quantum teleportation \cite{cst1,cst2}, and various quantum gates \cite{csqc,qg1}.\\
\indent Motivated by their usefulness and abandon applications in a large variety of disciplines \cite{uses5}, various generalization schemes have been introduced \cite{c2,c4} and the coherent states have been constructed for a large class of physical systems\cite{c4,gcs}. The generalized coherent states for the systems possessing non-linear energy spectrum may take the potential to describe the non-linear quantum optics and can be useful in related technologies\cite{c6}. Moreover, such constructions take the promise to converge various research areas to motivate new interdisciplinary research, for instance, most recently coherent states have been used in string theory \cite{stc}, squeezed states in non-commutative spaces \cite{ssnc} and their entanglement generation by means of beam splitters \cite{ebs}.\\
\indent In this article generalized coherent states have been discussed in the context of position-dependent effective mass (PDEM) systems. PDEM systems are of great interest due to vast applications in various areas of physics \cite{o1,prcgp,pn,com,nse,nss,nsnew,nsjmp}. The quantum mechanical description of such systems becomes challenging due to the existence of position dependence in the kinetic energy term. PDEM systems have been discussed extensively in the contexts of finding their solutions \cite{com}, ladder operators and associated algebras \cite{nsnew}, constructing the coherent states \cite{nss}.\\
\indent Most recently, a generalized scheme for constructing the ladder operator for PDEM systems have been introduced  \cite{nsnew}. In present work, we will use these ladder operators to construct the coherent states for PDEM systems. The general formalism has been applied to different quantum systems with spatially varying mass. Various properties of the coherent states for these systems has been discussed. We close our work by some concluding remarks.\\
\section{Ladder operators for PDEM systems}
\indent The classical dynamics of a PDEM system is governed by a Hamiltonian,  
\begin{equation}\label{1.1}
H=\frac{p^{2}}{2m(x)}+V(x,\alpha),
\end{equation}
which can be quantized by considering the symmetric ordering of the operators concerning momentum and spatially varying mass \cite{o1,nse,nss,nsnew,nsjmp} as
\begin{equation}\label{1.3}
\hat{H}=
-\bigg(\frac{1}{2m(x)}\bigg)\frac{d^{2}}{dx^{2}}-\bigg(\frac{1}{2m(x)}\bigg)^{'}\frac{d}{dx}+V(x,\alpha),
\end{equation}
where $\alpha$ represents the parameter that specify space-independent properties of the potential, such as range, strength and diffuseness.\\
\indent In order to obtain the corresponding ladder operators for the underlying system, we factorize the Hamiltonian given in Eq. (\ref{1.3}), as 
\[
\hat{H}=\hat{A}^{\dag}(\alpha)\hat{A}(\alpha)+E_{0},
\]
where $E_{0}$ is the ground-state energy of the Hamiltonian $\hat{H}$ and $\hat{A}(\alpha),~\hat{A}^{\dag}(\alpha)$ represent a pair of operators 
\begin{eqnarray}\label{op}
\hat{A}(\alpha)&=&\frac{1}{\sqrt{2m(x)}}\frac{d}{dx}+W(x,\alpha), \\
\hat{A^{\dag}}(\alpha)&=&\frac{-1}{\sqrt{2m(x)}}\frac{d}{dx}-\bigg(\frac{1}{\sqrt{2m(x)}}\bigg)^{'}+W(x,\alpha), \nonumber
\end{eqnarray}
where $W(x,\alpha)$ denotes the super-potential depending on the parameter $\alpha$ \cite{nsnew,g,b}. However, it is important to note that these operators can not be treated as the ladder operators since their commutator $[\hat{A}(\alpha),\hat{A}^{\dag}(\alpha)]$,
depends on the position variable $``x"$ \cite{nsnew}. In order to overcome this difficulty we need to introduce new operators whose commutator is independent of any dynamical variables. An integrability condition known as shape invariance (SI)  \cite{nsnew,nsjmp,g,b,f} plays a vital role in this regard. The SI condition in terms of the operators, defined in Eq. (\ref{op}), is given as 
\begin{equation}\label{4.1}
\hat{A}(\alpha_{1})\hat{A}^{\dag}(\alpha_{1})-\hat{A}^{\dag}(\alpha_{2})\hat{A}(\alpha_{2})=R(\alpha_{1}),
\end{equation}
where $\alpha_{1}=\alpha,~\alpha_{2}=\alpha_{1}+\eta,$ and $R(\alpha_{1})$ is the remainder term independent of the dynamical variables. This reparametrization of parameter $\alpha_{1}$ is achieved by means of a similarity transformation, 
\begin{equation}\label{ac}
\hat{T}^{-1}(\alpha_{1})R(\alpha_{n}) \hat{T}(\alpha_{1})= R(\alpha_{n-1}),
\end{equation}
where $\hat{T}(\alpha_{1})$ is a translation operator defined as \begin{equation}\label{st}
\hat{T}(\alpha_{1})|\varphi(\alpha_{1})\rangle=|\varphi(\alpha_{2})\rangle,
\end{equation}
such that $\hat{T}(\alpha_{1})\hat{T}^{-1}(\alpha_{1})=\mathbf{1}$. By means of the translation operator (\ref{st}) and the operators (\ref{op}), we introduce a pair of new operators
\begin{equation}\label{4.6}
\hat{L}_{-}(\alpha_{1})=\hat{T}^{-1}(\alpha_{1})\hat{A} (\alpha_{1}),~~
\hat{L}_{+}(\alpha_{1})=\hat{A}^{\dag} (\alpha_{1})\hat{T}(\alpha_{1}),
\end{equation}
so that the integrability condition (\ref{4.1}), together with the similarity transformation introduced in Eq. (\ref{ac}), takes the form
\begin{equation}
[\hat{L}_{-}(\alpha_{1}),~ \hat{L}_{+}(\alpha_{1})]=R(\alpha_{0}),
\end{equation} 
which resembles the well known Heisenberg Weyl algebra and this suggests us to consider $ \hat{L}_{\pm}(\alpha_{1})$ as the appropriate ladder operators \cite{nsnew}. These ladder operators satisfy the relation $R(\alpha_{n}) \hat{L}_{+}(\alpha_{1})= \hat{L}_{+}(\alpha_{1})R(\alpha_{n-1}),$ which provide us with the eigenvalues of $\hat{H}$ given in (\ref{1.3}), as
\begin{equation}\label{esg}
E_{n}=\sum_{k=1}^{n} R(\alpha_{k})+E_{0}.
\end{equation}
\indent In order to obtain the normalized eigenstates of $\hat{H}$, we see that the ladder operators act on the eigenstates $|\varphi_{n}(\alpha_{1})\rangle$, of the given system as
\begin{eqnarray}\label{2.19}
\hat{L}_{+}(\alpha_{1})|\varphi_{n}\rangle&=& \bigg[\sum_{k=1}^{n+1}R(\alpha_{k})\bigg]^{\frac{1}{2}}|\varphi_{n+1}\rangle,
\nonumber\\
\hat{L}_{-}(\alpha_{1})|\varphi_{n}\rangle&=& \bigg[\sum_{k=1}^{n}R(\alpha_{k})\bigg]^{\frac{1}{2}}|\varphi_{n-1}\rangle.
\end{eqnarray}
As a result the normalized eigenstates of $\hat{H}$ are given as
\begin{equation}\label{2.17}
|\varphi_{n}\rangle=\frac{1}{\sqrt{\rho_{n}}}~~[ \hat{L}_{+}(\alpha_{1})]^{n}|\varphi_{0}\rangle,
\end{equation}
where $\rho_{n}$ is the generalized factorial defined as
\begin{equation}\label{fac} 
\rho_{n}=[R(\alpha_{n})+R(\alpha_{n-1})+...+R(\alpha_{1})]\dots[R(\alpha_{1})].
\end{equation}
\section{Generalized coherent states}
\indent As mentioned before, the ladder operators provides a strong base for the construction of algebraic dependent coherent states. Earlier, this kind of states have been constructed for the constant mass systems \cite{G.1,f}. Our aim is to generalize this notion to incorporate the spatial dependence of mass. Assume that the systems under consideration have infinite bound states. Following the usual way of constructing coherent states for any quantum mechanical system, we define coherent states $|z\rangle$ as eigenstates of the lowering operator $\hat{L}_{-}$, introduced in Eq. (\ref{4.6}), as
\begin{equation}\label{5.1}
\hat{L}_{-}|z\rangle = z|z\rangle,
\end{equation}
where ``$z$'' is a complex parameter. In order to derive an explicit expression for these coherent states, we express $|z\rangle$ as a superposition of the eigenstates $|\varphi_{n}\rangle$ of the system under consideration as 
\begin{equation}\label{5.2}
|z\rangle=\sum_{n=0}^{\infty} c_{n}|\varphi_{n}\rangle.
\end{equation}
Using the above relation in Eq. (\ref{5.1}), we get
\begin{equation}
\sum_{n=0}^{\infty} c_{n}~ \hat{L}_{-}|\varphi_{n}\rangle=z~\sum_{n=0}^{\infty} c_{n}|\varphi_{n}\rangle,
\end{equation}
which on simplification provides us with the following equation
\begin{equation}
c_{n}=\frac{z^{n}}{\sqrt{\rho_{n}}}~c_{0},
\end{equation} 
where $c_{0}$ is a constant that needs to be determined and $\rho_{n}$ is the generalized factorial introduced in Eq. (\ref{fac}). Finally, Eq. (\ref{5.2}) can be rewritten as
\begin{equation}\label{5.4}
|z\rangle=c_{0}~\sum_{n=0}^{\infty} \frac{z^{n}}{\sqrt{\rho_{n}}} |\varphi_{n}\rangle.
\end{equation}
The unknown $c_{0}$ can be determined by using the normalization condition $\langle z|z\rangle=\mathbf{1}$, as
\begin{equation}
c_{0}=[\mathcal{N}(|z|^{2})]^{\frac{-1}{2}}=\bigg(\sum_{n=0}^{\infty}\frac{(|z|^{2})^{n}}{\rho_{n}}\bigg)^{\frac{-1}{2}}.
\end{equation}
Thus, the final form of the generalized coherent states for a quantum mechanical system with PDEM is given as
\begin{equation}\label{5.6}
|z\rangle=\frac{1}{\sqrt{\mathcal{N}(|z|^{2})}}\sum_{n=0}^{\infty} \frac{z^{n}}{\sqrt{\rho_{n}}} |\varphi_{n}\rangle.
\end{equation}
Note that the normalized coherent states defined above, satisfy the requirement of continuity of labeling as required for the coherent states \cite{klu.0}, since the transformation of coherent state parameters $z\rightarrow z^{'}$ leads to the transformation of the states $|z\rangle\rightarrow |z^{'}\rangle$. Another important observation about these states is that although the states $|z\rangle$ are normalized but they are not orthogonal to each other since 
\begin{equation}
\langle z|z^{'}\rangle=\frac{\mathcal{N}(z^{*}z^{'})}{\sqrt{\mathcal{N}(|z|^{2})~\mathcal{N}(|z^{'}|^{2})}}.\nonumber
\end{equation}\\
We now investigate the overcompleteness property of the generalized coherent states for the shape invariant potentials with PDEM. This property is commonly known as resolution of unity. For this we assume that there exist a positive and unique weight function $w(|z|^{2})$, such that
\begin{equation}\label{5.20}
\int d\mu |z\rangle\langle z|=\mathbf{1}=\sum_{n=0}^{\infty} |\varphi_{n}\rangle\langle \varphi_{n}|,
\end{equation}
where $d\mu=d^{2}z~w(|z|^{2})/\pi.$ Our aim is to determine this weight function. For this we use Eq. (\ref{5.6}) in the above equation and introduce the change of variables, $z=re^{i\theta},~~~|z|^{2}=r^{2},~~~d^{2}z=r dr d\theta.$ The angular integral leads to $\int_{0}^{2\pi} e^{i(n-m)\theta}d\theta=2\pi\delta_{n,m},$ so that our task of finding the weight function $w(|z|^{2})$, reduces to finding the solution of the radial integral equation $\int_{0}^{\infty}2rdr~ w(r^{2})\sum_{n=0}^{\infty} \frac{r^{2n}}{\rho_{n}~\mathcal{N}(r^{2})}  |\varphi_{n}\rangle\langle\varphi_{n}|=1,
$ which on introducing the change of variable $r^{2}=\xi$, takes the form, 
\begin{equation}\label{5.24}
\int_{0}^{\infty} \tilde{w}(\xi) \xi^{n} d\xi=\rho_{n},
\end{equation}
where we have used $\tilde{w}(\xi)=w(\xi)/\mathcal{N}(\xi).$ Note that (\ref{5.24}), is an inverse moment problem which can be solved by using well known Mellin transforms \cite{kps01} or by making use of the Meijer's G-function \cite{rkam}. We can also determine the correct form of the weight function $w(|z|^{2})$, by using Fourier transform technique.\\
\indent The radius of convergence for the coherent state $|z\rangle$, is defined as
\begin{equation}\label{5.25}
R=\lim_{n\rightarrow \infty}(\rho_{n})^{1/n}.
\end{equation}
This is important in the sense that any coherent state can only exist if the radius of convergence of that state is non-zero.\\
\indent The statistical features of any coherent state can be characterized by the probability distribution which is formally given as
\begin{equation}\label{wdg}
P_{n}=|\langle \varphi_{n}|z\rangle|^{2}=\frac{|z|^{2n}}{\rho_{n}~\mathcal{N}(|z|^{2})}.
\end{equation}
The mean and variance, which are used to characterize the weighting distribution of coherent states, can be calculated by using the first and second moments of the probability distribution. Mean corresponds to the first moment and is obtained as
\begin{equation}\label{mg}
\langle n \rangle=\frac{1}{\mathcal{N}(|z|^{2})}\sum_{n=0}^{\infty} \frac{n}{\rho_{n}}~|z|^{2n}.
\end{equation}
The second moments of the probability distribution is given as
\begin{equation}\label{smg}
\langle n^{2} \rangle=\frac{1}{\mathcal{N}(|z|^{2})}\sum_{n=0}^{\infty} \frac{n^{2}}{\rho_{n}}~|z|^{2n},
\end{equation}
so that the variance can be determined by the following relation
\begin{equation}\label{vg}
(\Delta n)^{2}= \langle n^{2} \rangle - \langle n \rangle ^{2}.
\end{equation}
The nature of weighting distribution of coherent states can be characterized by means of the Mandel's parameter \cite{gl.bk,mandel1} which is defined as
\begin{equation}\label{mpg}
Q=\frac{(\Delta n)^{2}}{\langle n\rangle}-1.
\end{equation}
The weighting distribution is Poissonian in nature if $Q = 0$, sub-Poissonian if $Q < 0$ and super-Poissonian for positive values of $Q$. \\
\section{Examples}
\begin{figure*}
	\centering
	\includegraphics[width=.48\textwidth]{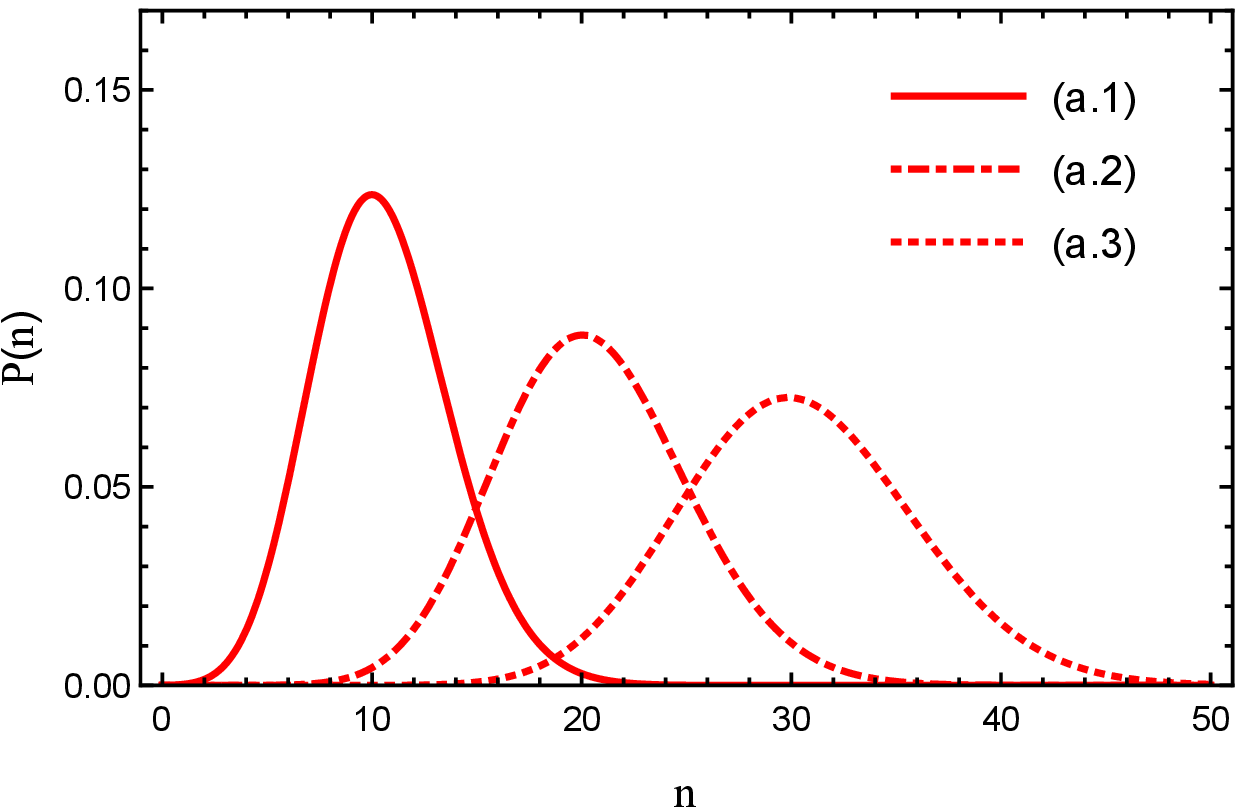}
		\includegraphics[width=.48\textwidth]{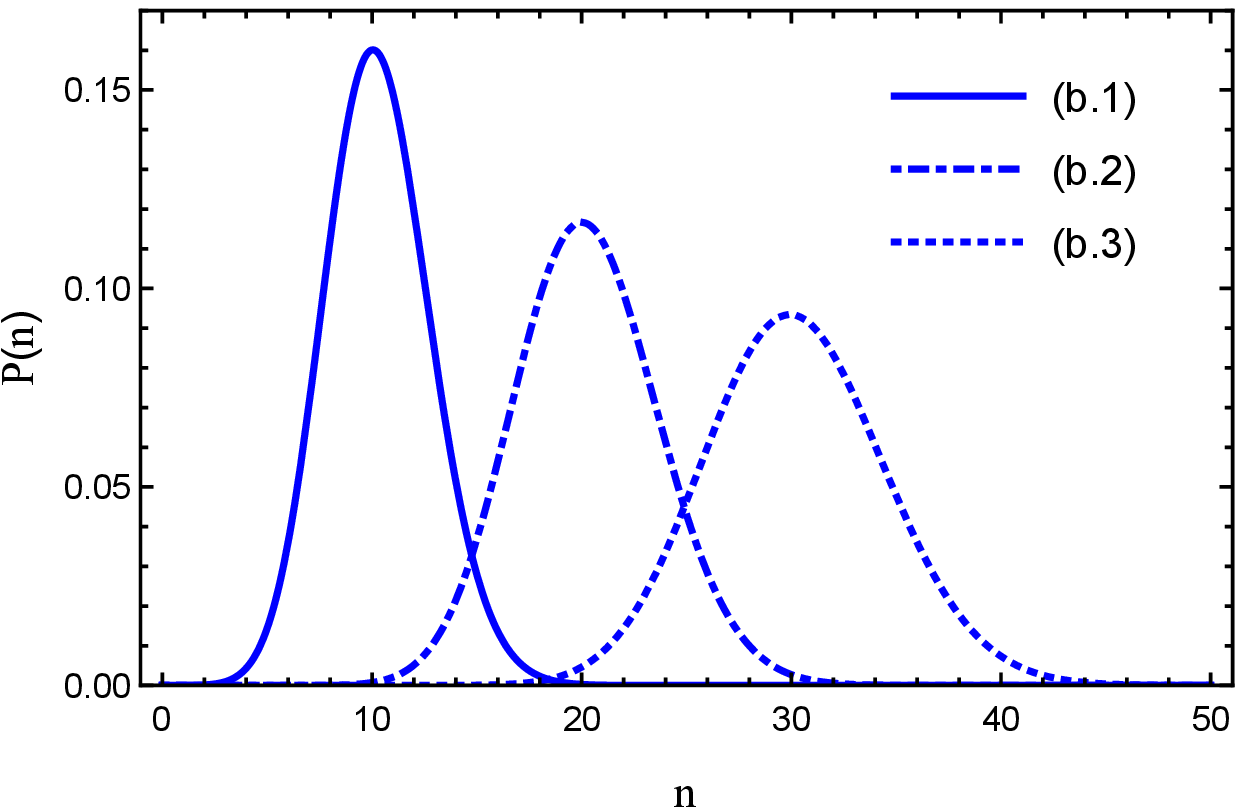}
	\caption{The weighting distribution $P_{n}$, defined in Eq. (\ref{wdho}), for the linear harmonic oscillator (left) and for the non-linear oscillator (right) as a function of quantum number $``n"$ for fixed different values of the nonlinearity parameter $``\lambda^{'}"$: (b.1) $\lambda^{'}=-0.27$, (b.2) $\lambda^{'}=-.17$, (b.3) $\lambda^{'}=-.07$.}\label{distho}
\end{figure*}

\indent In order to exemplify the general formalism presented in the previous section, we consider few PDEM systems with shape-invariant potentials. One case is presented in detail while for the sake of brevity we present main results for the remaining cases.\\
\indent \textit{\textbf{Example 1:}} Let us first consider a non-linear oscillator with potential $V(x)= m(x)\alpha^{2}x^{2}/2,$
where $m(x)=(1+\lambda x^{2})^{-1}$ and $\lambda$ is the non-linearity parameter. It is important to note that $\lambda$ can be positive as well as negative. However, for negative values of $\lambda$, there exists a singularity for the given mass function and associated dynamics, at $1-|\lambda|x^{2} = 0$. Therefore, for $\lambda<0$, our analysis is restricted to the interior of the interval $x^{2}\leq 1/|\lambda|$ \cite{nsjmp}. By using the symmetric ordering of $m(\hat{x})$ and $\hat{p}$ in the equivalent kinetic energy operator \cite{o1,nse,nss,nsjmp}, the quantized Hamiltonian is given as
\begin{equation}\label{h}
\hat{H}=\frac{\alpha}{2}\bigg[-(1+\tilde{\lambda} \zeta^{2})\frac{d^{2}}{d\zeta^{2}}-2\tilde{\lambda} \zeta\frac{d}{d\zeta}+\frac{\zeta^{2}}{1+\tilde{\lambda} \zeta^{2}}\bigg],
\end{equation}
where we have made use of the dimensionless variables $\zeta=\sqrt{\alpha}x$ and $\tilde{\lambda}=\lambda/\alpha$.\\
\indent It is important to note that for positive values of $\tilde{\lambda}$, we get a finite energy spectrum. However, for $\tilde{\lambda} < 0$, we have  $\tilde{\lambda}=-\mid\tilde{\lambda}\mid$ and the energy spectrum is unbounded. In this case there exists an infinite but countable eigenfunctions and for the upcoming analysis we shall consider this choice. Hence, for the present case, the eigenvalues, defined in (\ref{esg}), are given as 
\begin{equation}
E_{n} = \alpha \bigg( n+\frac{1}{2} + n(n+1)\mid\lambda^{'}\mid\bigg),~~n=0,1,2,3,\dots
\end{equation} 
where $\lambda^{'}=\mid\tilde{\lambda}\mid/2$. In this case the relations (\ref{2.19}), satisfied by the ladder operators are given as
\begin{eqnarray}\label{c43.10}
\hat{L}_{-}|\varphi_{n}\rangle &=& \sqrt{n+\lambda^{'}n(n+1)}
~~|\varphi_{n-1}\rangle, \\
\nonumber \hat{L}_{+}|\varphi_{n}\rangle &=& \sqrt{(n+1)+\lambda^{'}(n+1)(n+2)}
~~|\varphi_{n+1}\rangle.
\end{eqnarray}
With the help of these ladder operators, the eigenenergies and the corresponding eigenstates of the system under consideration, are respectively given as
\begin{equation}\label{c43.11}
|\varphi_{n}\rangle= \frac{[\hat{L}_{+}]^{n}|\varphi_{0}\rangle}{\sqrt{\rho_{n}}},
\end{equation}
where 
\begin{eqnarray}\label{facho}
\rho_{n}=\frac{n!~(\lambda^{'})^{n}\Gamma\big(2+\frac{1}{\lambda^{'}}+n\big)}{\Gamma\big(2+\frac{1}{\lambda^{'}}\big)},~~\rho_{0}=1.
\end{eqnarray}
\indent In order to define the coherent states of the non-linear oscillator with PDEM as the eigenstates of the lowering operator, we consider Eq. (\ref{5.6}), which on using Eq. (\ref{facho}), becomes
\begin{equation}\label{e5.1}
|z\rangle=\frac{1}{\sqrt{\mathcal{N}(|z|^{2})}}
\sum_{n=0}^{\infty} \bigg[\frac{\Gamma\big(2+\frac{1}{\lambda^{'}}\big)}{n!~\Gamma\big(2+\frac{1}{\lambda^{'}}+n\big)}
\bigg(\frac{1}{\lambda^{'}}\bigg)^{n}\bigg]^{\frac{1}{2}} z^{n}|\varphi_{n}\rangle,
\end{equation}
where $\mathcal{N}(|z|^{2})=~_{0}F_{1}\bigg(2+\frac{1}{\lambda^{'}};\frac{|z|^{2}}{\lambda^{'}}\bigg).$ Note that the coherent states defined in Eq. (\ref{e5.1}), satisfies the Klauder's minimal set of conditions that are required for any coherent state \cite{klu.0}. The overlap of two coherent states for the non-linear oscillator is given as 
\begin{equation}
\langle z|z^{'} \rangle=\frac{1}{\sqrt{ N(|z|^{2}) N(|z^{'}|^{2})}}
~_{0}F_{1}\bigg(2+\frac{1}{\lambda^{'}};\frac{z^{'} z^{*}}{\lambda^{'}} \bigg),\nonumber
\end{equation}
from which it follows that the coherent states for the non-linear oscillator with PDEM are not orthogonal. The continuity in the label $z$ follows immediately because of the fact that 
\begin{equation}
\lim_{z^{'}\rightarrow z} \parallel |z^{'}\rangle-|z\rangle\parallel^{2}=
\lim_{z^{'}\rightarrow z} [2(1-Re \langle z^{'}|z\rangle)]=0.
\end{equation}
We now investigate the over-completeness of the coherent states defined in Eq. (\ref{e5.1}). We look for a positive and unique weight function $w(|z|^{2})$, such that Eq. (\ref{5.20}) is satisfied. Substituting (\ref{e5.1}) in (\ref{5.20}) and introducing the variables $z=re^{i\theta}$ and $|z|^{2}=\xi$, we finally arrive at
\begin{equation}\label{e5.5}
\int_{0}^{\infty}  \tilde{w}(\xi)\xi^{n} d\xi = \frac{\Gamma(n+1)\Gamma(2+\frac{1}{\lambda^{'}}+n)}
{\Gamma(2-\frac{1}{\lambda^{'}})}(\lambda^{'})^{n},
\end{equation}
where $\tilde{w}(\xi)=w(\xi)/N(\xi)$. The weight function can be determined by using inverse Mellin transform. By using the Mellin transform of the Meijer's $G$-function \cite{rkam}, the required weight function $w(\xi)$, can be obtained as
\begin{equation}\label{e5.7}
w(\xi)=\frac{~_{0}F_{1}\bigg(2+\frac{1}{\lambda^{'}};\frac{\xi}{\lambda^{'}}\bigg)}
{\lambda^{'}\Gamma(2+\frac{1}{\lambda^{'}})}
G_{0,2}^{2,0}\bigg(  \begin{array}{c}
.... \\
0,1+\frac{1}{\lambda^{'}}
\end{array}\bigg| \frac{\xi}{\lambda^{'}}
\bigg),
\end{equation}
which satisfies the integral equation (\ref{e5.5}). By using (\ref{5.25}), the radius of convergence for the non-linear oscillator with PDEM is given as
\begin{equation}\label{e5.8}
R=\lim_{n\rightarrow \infty} \bigg[n!(\lambda^{'})^{n}\bigg(2+\frac{1}{{\lambda}^{'}}\bigg)_{n}\bigg]^{\frac{1}{n}}
=\infty.
\end{equation}
This shows that the coherent states for the non-linear oscillator with PDEM are defined on the whole complex plane.\\
\indent Now we examine statistical properties of the non-linear oscillator with spatially varying mass. The probability distribution of the non-linear oscillator with PDEM for the generalized coherent states turns out to be
\begin{equation}\label{wdho}
P_{n}=\frac{1}{\mathcal{N}(|z|^{2})}
\bigg[\frac{\Gamma\big(2+\frac{1}{\lambda^{'}}\big)}{n!\Gamma\big(2+\frac{1}{\lambda^{'}}+n\big)}\bigg]
\bigg(\frac{|z|^{2}}{\lambda^{'}}\bigg)^{n},
\end{equation}
which is plotted, in Fig. (\ref{distho}).
It is clear from the figure that the distribution for the non-linear oscillator with PDEM is narrower than the weighting distribution for the linear Harmonic oscillator which clearly indicates the sub-Poissonian nature of the distribution. Note that for the harmonic limit as the non-linearity parameter $\lambda^{'}$ approaches to zero the sub-Poissonian nature of the distribution tends to Poissonian one. \\
\indent The first moment of the weighting distribution of coherent states, which corresponds to the mean is calculated as
\begin{equation}\label{m}
\langle n \rangle=\frac{|z|^{2}}{(1+2\lambda^{'})}
\frac{~_{0}F_{1}\bigg(3+\frac{1}{\lambda^{'}};\frac{|z|^{2}}{\lambda^{'}}\bigg)}
{~_{0}F_{1}\bigg(2+\frac{1}{\lambda^{'}};\frac{|z|^{2}}{\lambda^{'}}\bigg)},
\end{equation}
and the second moments of the probability distribution is given as
\begin{equation}
\begin{small}\langle n^{2} \rangle=\langle n \rangle+\frac{|z|^{4}}{(1+2\lambda^{'})(1+3\lambda^{'})}
\frac{~_{0}F_{1}\bigg(4+\frac{1}{\lambda^{'}};\frac{|z|^{2}}{\lambda^{'}}\bigg)}
{~_{0}F_{1}\bigg(2+\frac{1}{\lambda^{'}};\frac{|z|^{2}}{\lambda^{'}}\bigg)},\nonumber
\end{small}
\end{equation}
so that the variance of the generalized coherent states comes out to be
\begin{equation}\label{v}
(\Delta n)^{2}= \langle n \rangle[1-\langle n \rangle]+\frac{|z|^{4}~_{0}F_{1}\bigg(4+\frac{1}{\lambda^{'}};\frac{|z|^{2}}{\lambda^{'}}\bigg)}{(1+2\lambda^{'})(1+3\lambda^{'})\mathcal{N}(|z|^{2})}.
\end{equation}
The Mandel's parameter defined in Eq. (\ref{mpg}), is given as
{\small
\begin{equation}
Q=\frac{|z|^{2} ~_{0}F_{1}\bigg(4+\frac{1}{\lambda^{'}};\frac{|z|^{2}}{\lambda^{'}}\bigg)}
	                  {(1+3\lambda^{'})~_{0}F_{1}\bigg(3+\frac{1}{\lambda^{'}};\frac{|z|^{2}}{\lambda^{'}}\bigg)}
                	-
                 \frac{|z|^{2} ~_{0}F_{1}\bigg(3+\frac{1}{\lambda^{'}};\frac{|z|^{2}}{\lambda^{'}}\bigg)}{(1+2\lambda^{'})\mathcal{N}(|z|^{2})}.	\nonumber 
\end{equation}
}
This clearly indicates the sub-Poissonian nature of the weighting distribution.\\
\indent \textit{\textbf{Example 2:}} Let us now consider a harmonic potential $V(x)= m(x)\alpha^{2}x^{2}/2,$ with mass profile $m(x)=(1-(\lambda x)^{2})^{-1}.$  The mass profile encounters a singularity for both positive and negative values of $\lambda$ and our study of dynamics is restricted to the interior of the interval $x^{2}\leq 1/\lambda^{2}$ \cite{nsjmp}.\\
\indent For the present case, the energy spectrum is given as
\begin{equation}
E_{n}=\alpha \bigg( n+\frac{1}{2}+\frac{\upsilon^{2}}{2}n(n+1)\bigg),
\end{equation}
where $\varrho=x\sqrt{2\alpha}$ and $\upsilon=\lambda/\sqrt{2\alpha}$, are the dimensionless variables. The ladder operators $\hat{L}_{\pm}$, satisfy the following relations
\begin{eqnarray}
\hat{L}_{-}|\varphi_{n}\rangle &=& \sqrt{n+\frac{\upsilon^{2}}{2}n(n+1)}
~~|\varphi_{n-1}\rangle, \\
\nonumber \hat{L}_{+}|\varphi_{n}\rangle &=& \sqrt{(n+1)+\frac{\upsilon^{2}}{2}(n+1)(n+2)}
~~|\varphi_{n+1}\rangle.
\end{eqnarray}
By means of the ladder operators, the normalized eigenstates of the underlying system can be obtained as
\begin{equation}
|\varphi_{n}\rangle= \frac{[\hat{L}_{+}]^{n}|\varphi_{0}\rangle}
{\sqrt{\rho_{n}}},{\ \ \ } \mbox{where}{\ \ \ } \rho_{n}=\bigg(\frac{\upsilon^{2}}{2}\bigg)^{n}n!\bigg(2+\frac{2}{\upsilon^{2}}\bigg)_{n}.\nonumber
\end{equation}
For this particular system the coherent states takes the form
\begin{equation}
|z\rangle=\frac{1}{\sqrt{\mathcal{N}(|z|^{2})}}
\sum_{n=0}^{\infty} \bigg[\frac{\Gamma\big(2+\frac{2}{\upsilon^{2}}\big)}{n!~\Gamma\big(2+\frac{2}{\upsilon^{2}}+n\big)}\bigg]^{\frac{1}{2}}
\bigg(\frac{\sqrt{2}z}{\upsilon}\bigg)^{n}|\varphi_{n}\rangle,
\end{equation}
where $\mathcal{N}(|z|^{2})=~_{0}F_{1}\bigg(2+\frac{2}{\upsilon^{2}}
	;\frac{2|z|^{2}}{\upsilon^{2}}\bigg).$
One can easily verify that these coherent states satisfy the Klauder's minimal set of conditions that are required for any coherent state \cite{klu.0}. \\
\indent \textit{\textbf{Example 3:}} Let us now consider the potential of the form $V(x,\alpha)=[\mu^{2}m(x)\{(\alpha^{2}-1)e^{2\mu x}+1\}-\mu^{2}(\alpha+1)]/2,$
with PDEM $m(x)=e^{-\mu x}/2,~~\mu>0$.
For the present case,
the ladder operators satisfy the following relations \cite{nsnew}
\begin{eqnarray}
\hat{L}_{-}|\phi_{n}\rangle=\mu\sqrt{n}|\phi_{n-1}\rangle,~~
\hat{L}_{+}|\phi_{n}\rangle=\mu\sqrt{n+1}|\phi_{n+1}\rangle.
\end{eqnarray}
The energy spectrum and the corresponding eigenstates in this case turn out to be
\begin{equation}
E_{n}=n\mu^{2}~~\mbox{and}~~|\varphi_{n}\rangle= \frac{1}{\sqrt{n!}}
\bigg(\frac{\hat{L}_{+}}{\mu^{2}}\bigg)^{n}|\varphi_{0}\rangle,
\end{equation}
respectively.\\
\indent By using the above information in Eq. (\ref{5.6}), we get the coherent states for the system under consideration as
\begin{equation}\label{csm}
|z\rangle=\frac{1}{\sqrt{\mathcal{N}(|z|^{2})}}
\sum_{n=0}^{\infty} \frac{1}{\sqrt{n!}}\bigg(\frac{z}{\mu}\bigg)^{n}|\varphi_{n}\rangle,
\end{equation}
where $\mathcal{N}(|z|^{2})=e^{\big(\frac{|z|}{\mu}\big)^{2}}$ is the normalization constant. One can easily show that these states satisfies the Klauder's minimal set of conditions that are required for any coherent state \cite{klu.0}. For the sake of brevity we just compute the resolution of identity. As suggested in Eq. (\ref{5.20}), our aim is to determine a positive and unique weight function $w(|z|^{2})$. 
For the present case, the required weight function $w(\zeta)=\mu^{-2}$.
The radius of convergence for the pertaining system is given as $R=\lim_{n\rightarrow \infty}(n!\mu^{2n})^{\frac{1}{n}}=\infty,$
which shows that the coherent states for the present case are defined on the entire complex plane.\\
\indent The weighting distribution (\ref{wdg}), in this case turns out to be
\begin{equation}\label{wdm}
P_{n}=\frac{e^{-\big(\frac{|z|}{\mu}\big)^{2}}}{n!}\bigg(\frac{|z|}{\mu}\bigg)^{2n}.
\end{equation}
The mean and variance of the weighting distribution are given by
\begin{equation}\label{mm}
\langle n \rangle=\bigg(\frac{|z|}{\mu}\bigg)^{2}
~~\mbox{and}~~(\Delta n)^{2}= \bigg(\frac{|z|}{\mu}\bigg)^{2}.
\end{equation}
Since mean and variance are equal in this case, therefore, it clearly indicates the Poissonian nature of the distribution. Moreover, by making use of Eqs. (\ref{mm}) in Eq. (\ref{mpg}), we get $Q=0$, which is property of the standard harmonic oscillator.
\section{Summary and Conclusions}
A general scheme for constructing coherent states, in the context of position-dependent effective mass systems, have been discussed by means of the ladder operators and associated algebra of the system. An integrability condition , namely translational shape-invariance, has been used to find the ladder operators, energy spectrum and associated algebra for the position-dependent effective mass systems. For the constructed coherent states various properties have been analyzed. In order to illustrate the general formalism, we considered several shape-invariant potentials with position-dependent effective mass. The work is entirely interdisciplinary which may lead to various research areas, such as, condensed matter physics, quantum optics and quantum information theory to initiate new directions of research.


\begin{thebibliography}{00}
\bibitem{c5}   Schr\"{o}dinger E., \textit{Naturwissenschaften} {\bf 14} (1926) 664.

\bibitem{G.1}  Glauber R. J., \textit{Phys. Rev. Lett.} {\bf 10} (1963) 277; Glauber R. J., \textit{ Phys. Rev. {\bf 130}} (1963) 2529; Glauber R. J., \textit{Phys. Rev. {\bf 131}} (1963) 2766.

\bibitem{gl.bk} Glauber R. J., \textit{Quantum theory of optical coherence: selected papers and lectures} (John Wiley $\&$ Sons) 2007.

\bibitem{csqi1}  Cerf N. J.,  Leuchs G. and  Polzik E. S., \textit{Quantum Information with Continuous Variables of Atoms and Light} (Imperial College Press) 2007; Andersen U. L., Leuchs G. and Silberhorn C., \textit{Laser \& Photon. Rev.} \textbf{4} (2010) 337.

\bibitem{csqc}  Ralph T. C., Gilchrist A., Milburn G. J., Munro W. J. and Glancy S., \textit{Phys. Rev. A} \textbf{68} (2003) 042319. 

\bibitem{qm}   Joo J., Munro W. J. and Spiller T. P., \textit{Phys. Rev. Lett.} \textbf{107} (2011) 083601.
 
\bibitem{cst1}  Wang X., \textit{Phys. Rev. A} \textbf{64} (2001) 022303.

\bibitem{cst2}   van Enk S. J. and Hirota O., \textit{Phys. Rev. A} \textbf{64} (2001) 022313.

\bibitem{qg1}    Marek P. and Fiur\'{a}sek J., \textit{Phys. Rev. A} \textbf{82} (2010) 014304.

\bibitem{uses5}  Klauder J. R. and Skagerstam B., \textit{Coherent States: Applications in Physics and Mathematical Physics} (World Scientific) 1985;
                 Ali S. T., Antoine J. P. and Gazeau J. P., \textit{Coherent States, Wavelets and Their Generalizations} (Springer-Verlag, New York) 2000. 

\bibitem{c2}     Zhang W. M., Feng D.H. and  Gilmore R., \textit{ Rev. Mod. Phys.} \textbf{62} (1990) 867; Klauder J. R., \textit{ Phys. Rev. D} \textbf{19} (1979) 2349.

\bibitem{c4}    Perelomov A., \textit{Generalized Coherent States and Their Applications} (Springer-Verlag, Heidelberg) 1986;\\
                Perelomov A., \textit{Commun. Math. Phys}. \textbf{26} (1972) 222.

\bibitem{gcs}  Iqbal S., Riviere P. and Saif F., \textit{Int. J. Theor. Phys.} \textbf{49} (2010) 2540;
			  Iqbal S. and Saif F., \textit{J. Math. Phys.} \textbf{52} (2011) 082105;
              Iqbal S. and Saif F., \textit{Phys. Lett. A} \textbf{376} (2012) 1531;
              Iqbal S. and Saif F., \textit{J. Russ. Laser Res.} \textbf{34} (2013) 77.

\bibitem{barut1971new}  Barut A. and Girardello L., \textit{Commun. Math. Phys.} \textbf{21} (1971) 41.

\bibitem{c6}  Dodonov V. V., J. \textit{Opt. B} \textbf{4} (2002) R1;
             Dodonov V. V and Manko V. I., \textit{Theory of nonclassical states of light} (Taylor and Francis, New York) 2003;
              Nieto L. M.,\textit{ AIP Conf. Proc.} \textbf{809} (2006) 3.

\bibitem{stc}  Iqbal S., \textit{Phys. Lett. B} \textbf{725} (2013) 487;
               Ghosh S. and Roy P., \textit{Phys. Lett. B} \textbf{711} (2012) 423.

\bibitem{ssnc}  Dey S. and Fring A., \textit{Phys. Rev. D} \textbf{86} (2012) 064038.

\bibitem{ebs}  Ralph T. C., Gilchrist A., Milburn G. J., Munro W. J. and Glancy S., \textit{Phys. Rev. A} \textbf{68} (2003) 042319.

\bibitem{o1}   Von Roos O, \textit{Phys. Rev. B} \textbf{27} (1983) 7547; Von Roos O. and Mavromatis H., \textit{Phys. Rev. B} \textbf{31} (1985) 2294; Geller M. R. and Kohn W., \textit{Phys. Rev. Lett.} \textbf{70} (1993) 3103; De Saavedra F. A., Boronat J., Polls A. and Fabrocini A., \textit{Phys. Rev. B} \textbf{50} (1994) 4248; L\'{e}vy-Leblond J. M., \textit{Phys. Rev. A} \textbf{52} (1995) 1845; Barranco M., Pi M., Gatica S. M., Hernandez E. S. and Navarro J., \textit{Phys. Rev. B} \textbf{56} (1997) 8997.

\bibitem{prcgp}   Plastino A. R., Rigo A., Casas M., Garcias F. and Plastino A., \textit{Phys. Rev. A} \textbf{60} (1999) 4318.
\bibitem{pn}    A. J. Peter, K. Navaneethakrishnan, \textit{Physica E} \textbf{40} (2008) 2747.
\bibitem{nse}   Amir N. and Iqbal S., \textit{J. Math. Phys.} \textbf{55} (2014) 0114101.
\bibitem{com}   Amir N. and Iqbal S., \textit{Commun. Theor. Phys.} \textbf{62} (2014) 790.
\bibitem{nsnew}  Amir N. and Iqbal S., \textit{EPL} \textbf{111} (2015) 20005.
\bibitem{nss}   Amir N. and Iqbal S., \textit{J. Math. Phys.} \textbf{56} (2015) 062108.
\bibitem{nsjmp} Amir N. and Iqbal S., \textit{J. Math. Phys.} \textbf{57} (2016).
\bibitem{g}     Gendenshtein L., \textit{Pisma Z. Eksp. Teor. Fiz.} \textbf{38} (1983) 299
(Engl. trans. \textit{JETP Lett.} \textbf{38} (1983) 356).
\bibitem{b}     Balantekin A. B., \textit{Phys. Rev. A} \textbf{57} (1998) 4188.
\bibitem{f}     Fukui T. and Aizawa N., \textit{Phys. lett. A} \textbf{180} (1993) 308; Balantekin A. B., Ribeiro M. A. C. and Aleixo A. N. F., \textit{J. Phys. A: Math. Gen.} \textbf{32} (1999) 2785; Aleixo A. N. F. and Balantekin A. B., \textit{J. Phys. A: Math. Gen.} \textbf{37} (2004) 8513.
\bibitem{klu.0} Klauder J. R., \textit{J. Math. Phys.} \textbf{4} (1963) 1055.
\bibitem{kps01}  Klauder J. R., Penson K. A. and Sixdeniers J. M., \textit{Phys. Rev. A} \textbf{64} (2001) 013817.
\bibitem{rkam}  Mathai A. M. and Saxena R. K., \textit{Generalized hypergeometric functions with applications in statistics and physical sciences} (Volume 348, Springer) 1973.
\bibitem{mandel1} Mandel L, \textit{Optics Letters} \textbf{4} (1979) 205; Mandel L. and Wolf E., \textit{Optical coherence and quantum optics} (Cambridge University Press) 1995.
\end{thebibliography}
\end{document}